\documentclass[%
 reprint,
 amsmath,amssymb,
 aps,
showkeys
]{revtex4-1}

\usepackage{graphicx}
\usepackage{dcolumn}
\usepackage{bm}

\usepackage{xcolor}

\begin{document}


\title{A novel approach for deriving the stable boundary layer height and\\ eddy viscosity profiles from the Ekman equations}

\author{Sukanta Basu}
\email{sukanta.basu@gmail.com}
\affiliation{Faculty of Civil Engineering and Geosciences, Delft University of Technology, Delft, the Netherlands
}
\author{Albert A. M. Holtslag}
\email{bert.holtslag@wur.nl}
\affiliation{Meteorology and Air Quality, Wageningen University, Wageningen, the Netherlands}%

\date{\today}

\begin{abstract}
In this study, we utilize a novel approach to solve the Ekman equations for eddy viscosity profiles in the stable boundary layer. By doing so, a well-known expression for the stable boundary layer height by Zilitinkevich (Boundary-Layer Meteorology, 1972, Vol. 3, 141--145) is rediscovered.
\end{abstract}


\maketitle

\section{Introduction}
\label{intro}

Sergej Zilitinkevich was one of the giants of atmospheric physics who carried the field of boundary-layer meteorology (BLM) on his shoulder for more than half a century. We, representing the BLM community at large, are indebted forever for his ingenious efforts and lifelong dedication in advancing our field. In the arena of stable boundary layers (SBLs), Zilitinkevich made numerous ground-breaking contributions. As a matter of fact, it would be difficult to find any contemporary article on SBLs which does not make at least one reference to an original publication of Zilitinkevich. The present paper is also following the same tradition. 

In this study, we utilize the Ekman equations to analytically derive a stable boundary layer height formulation which was originally proposed by Zilitinkevich~\cite{zilitinkevich72} based on scaling arguments. During the process, we also derive equations for the vertical profiles of eddy viscosity. To the best of our knowledge, it is the first time that estimates are given for the eddy viscosity profiles directly from the Ekman equations. 

\section{Formulation of SBL Height by Zilitinkevich (1972)}
\label{Z72}

Using boundary layer scaling arguments, Zilitinkevich~\cite{zilitinkevich72} proposed that the height ($h$) of stationary SBLs can be written as: 

\begin{equation}
    h = \gamma \sqrt{\frac{u_{*0} L}{f_{cor}}} = C_h u_{*0}^2 \left| f_{cor} B_s \right| ^{-1/2}.
    \label{EqZ72}
\end{equation}
Here the surface friction velocity and surface Obukhov length are denoted by $u_{*0}$ and $L$, respectively; $B_s$ is the buoyancy flux. The Coriolis term is represented by $f_{cor}$. The proportionality constants $\gamma$ and $C_h$ are related as follows: 
\begin{equation}
    C_h = \frac{\gamma}{\sqrt{\kappa}} \approx 1.58 \gamma,
\end{equation}
where $\kappa$ is the von K\'{a}rm\'{a}n constant assumed to be equal to 0.4. 
Zilitinkevich~\cite{zilitinkevich72} assumed $C_h$ to be order of one. In an analytical study, Businger and Arya~\cite{businger74} estimated $\gamma$ to be equal to 0.72. A much lower value of $\gamma \approx 0.4$ was estimated by Brost and Wyngaard~\cite{brost78} using a second-order closure model. Garratt~\cite{garratt82} who analyzed observational data from several field campaigns also found $\gamma \approx 0.4$. In his local-scaling paper, Nieuwstadt~\cite{nieuwstadt84} found $\gamma \approx 0.35$ to be consistent with other equations. Zilitinkevich~\cite{zilitinkevich89} summarized $C_h$ from several studies and found it to vary within the range of 0.55--1.58. A unique aspect of the present study is that we analytically derive $\gamma$ (and $C_h$) with limited assumptions. 

We would like to point out that Eq.~\ref{EqZ72} predicts physically unrealistic boundary layer heights for two situations: (i) close to the equator (i.e., $f_{cor} \to 0$); and (ii) for near-neutral conditions (i.e., $L \to \infty$). To circumvent the second issue, a few interpolation approaches have been proposed in the literature \citep[e.g.,][]{nieuwstadt81,zilitinkevich89}.

\section{Momentum and Sensible Heat Flux Profiles}

In SBL flows, the profiles of friction velocity and sensible heat flux are often expressed as follows:  
\begin{subequations}
\begin{equation}
u_{*L}^2 = \left(\overline{uw}_L^2 + \overline{vw}_L^2\right)^{1/2} =  u_{*0}^2 \left( 1 - \frac{z}{h} \right)^{\alpha} = u_{*0}^2 f_m,    
\label{ustar}
\end{equation}
\begin{equation}
\overline{w\theta}_L =  \overline{w\theta}_0 \left( 1- \frac{z}{h} \right)^{\beta} = \overline{w\theta}_0 f_h.
\label{SHFX}
\end{equation}
\end{subequations}
Here the friction velocity, momentum flux components, and sensible heat flux at height $z$ are denoted by $u_{*L}$, $\overline{uw}_L$, $\overline{vw}_L$, and  $\overline{w\theta}_L$,  respectively. The subscript `0' is used to demarcate the corresponding surface values. These equations imply that the turbulent fluxes are maximum near the surface and they monotonically decrease to zero at the top of the boundary layer. 

The exponents $\alpha$ and $\beta$ in Eqs.~\ref{ustar} and \ref{SHFX} are not universal constants. By utilizing his local-scaling hypothesis,  Nieuwstadt~\cite{nieuwstadt84} suggested $\alpha = 1.5$. He also proved that $\beta$ should be equal to 1 under the assumptions of horizontal homogeneity and stationarity. In contrast,  based on observational data from the well-known Minnesota field campaign \citep{caughey79}, Sorbjan~\cite{sorbjan86} estimated $\alpha = 2$,  and $\beta = 3$ for evolving SBLs.  In another observational study,  Lenschow et al.~\cite{lenschow88} considered the additional effects of radiational cooling and found the optimal $\alpha$ and $\beta$ to be equal to 1.75 and 1.5,  respectively. 

Using the definition of Obukhov length \citep{stull88} in conjunction with Eqs.~\ref{ustar} and \ref{SHFX}, we get: 
\begin{equation}
\Lambda = L \frac{f_m^{3/2}}{f_h},
\label{Obukhov}
\end{equation}
where $\Lambda$ is the so-called local Obukhov length at height $z$. 

The first and second derivatives of $f_m$ function in Eq.~\ref{ustar} can be written as:
\begin{subequations}
\begin{equation}
f_m' = - \frac{\alpha}{h} \left(1 - \frac{z}{h}\right)^{\alpha-1},
\label{Eqdfmdz}
\end{equation}
\begin{equation}
f_m'' = \frac{\alpha\left(\alpha - 1\right)}{h^2} \left(1 - \frac{z}{h}\right)^{\alpha-2}.
\label{Eqd2fmdz2}    
\end{equation}
\end{subequations}
We will make use of these derivatives in a later section. 

\section{Eddy Viscosity Profile in the Ekman Layer}
\label{EddyVisc}

According to the K-theory,  based on the celebrated hypothesis of Boussinesq in 1877, turbulent fluxes can be approximated as products of the eddy exchange coefficients and the mean gradients \citep{lumley64}. 

\begin{subequations}
\begin{equation}
\overline{uw}_L = - K_M \frac{\partial U}{\partial z},
\label{EqKm1}
\end{equation}
\begin{equation}
\overline{vw}_L = - K_M \frac{\partial V}{\partial z}.   
\label{EqKm2}
\end{equation}
\end{subequations}
Here $K_M$ is the so-called eddy viscosity coefficient. Based on these equations, we can write: 
\begin{equation}
    u_{*L}^2 = \left(\overline{uw}_L^2 + \overline{vw}_L^2\right)^{1/2} = K_M S,
    \label{EqKm3}
\end{equation}
where $S$ is the magnitude of wind speed shear. 

For boundary layer flows over homogeneous and flat terrain, under steady-state conditions, the averaged equations of motions can be simplified as follows \citep{zilitinkevich67,brown74}:

\begin{subequations}
\begin{equation}
f_{cor} \left(V - V_g\right) = \frac{\partial \left(\overline{uw}_L\right)}{\partial z},
\label{EqEkman1}
\end{equation}
\begin{equation}
f_{cor} \left(U - U_g\right) = - \frac{\partial \left(\overline{vw}_L\right)}{\partial z}.
\label{EqEkman2}
\end{equation}
\end{subequations}
The velocity components in $x$ and $y$ directions are represented as $U$ and $V$, respectively. Similarly, the geostrophic velocity components are denoted by $U_g$ and $V_g$. In the literature, these equations are commonly known as the Ekman layer equations \citep{brown74,nieuwstadt83}. 

In a landmark paper, Ekman~\cite{ekman05} first analytically solved  Eqs.~\ref{EqEkman1} and \ref{EqEkman2} with the assumption of constant eddy viscosity ($K_M$) in conjunction with appropriate boundary conditions. Over the years, a few more closed form analytical solutions of the Ekman equations have been reported in the literature \citep[e.g.,][]{wippermann73,brown74,nieuwstadt83,grisogono95,parmhed05}. In all these papers, simplified profiles of $K_M$ were always assumed. In this paper, we take a radically different approach. We only assume that Eq.~\ref{ustar} is valid and then deduce $K_M$ profile from the Ekman equations as shown below.   

For barotropic condition, the Ekman equations can be re-written as:
\begin{subequations}
\begin{equation}
f_{cor} \frac{\partial V}{\partial z} = \frac{\partial^2 \left(\overline{uw}_L\right)}{\partial z^2},
\label{EqEkman3}
\end{equation}
\begin{equation}
f_{cor} \frac{\partial U}{\partial z} = - \frac{\partial^2 \left(\overline{vw}_L\right)}{\partial z^2}.   
\label{EqEkman4}
\end{equation}
\end{subequations}
Next, by utilizing Eqs.~\ref{EqKm1} and \ref{EqKm2}, these equations are transformed as follows: 
\begin{subequations}
\begin{equation}
- f_{cor} \frac{\overline{vw}_L}{K_M} = \frac{\partial^2 \left(\overline{uw}_L\right)}{\partial z^2},   
\label{EqEkman5}
\end{equation}
\begin{equation}
- f_{cor} \frac{\overline{uw}_L}{K_M}  = - \frac{\partial^2 \left(\overline{vw}_L\right)}{\partial z^2}.    
\label{EqEkman6}
\end{equation}
\end{subequations}
Dividing Eq.~\ref{EqEkman5} by Eq.~\ref{EqEkman6} and rearranging we arrive at:
\begin{equation}
\overline{uw}_L \frac{\partial^2 \left(\overline{uw}_L\right)}{\partial z^2} + \overline{vw}_L \frac{\partial^2 \left(\overline{vw}_L\right)}{\partial z^2} = 0.
\label{EqEkmanCombined1}
\end{equation}
The momentum flux components can be decomposed in terms of local friction velocity as follows:
\begin{subequations}
\begin{equation}
\overline{uw}_L = - u_{*L}^2 \cos\left(\delta \right), 
\label{EqUW1}
\end{equation}
\begin{equation}
\overline{vw}_L = u_{*L}^2 \sin\left(\delta \right).    
\label{EqVW1}
\end{equation}
\end{subequations}
Here $\delta$ is the angle between the flux vector and the $x$-axis. By plugging Eq.~\ref{ustar} in Eqs.~\ref{EqUW1} and \ref{EqVW1}, we get: 
\begin{subequations}
\begin{equation}
\overline{uw}_L = - u_{*0}^2 f_m \cos\left(\delta \right),
\label{EqUW2}
\end{equation}
\begin{equation}
\overline{vw}_L = u_{*0}^2 f_m \sin\left(\delta \right).    
\label{EqVW2}
\end{equation}
\end{subequations}
These equations can be differentiated as follows:
\begin{subequations}
\begin{equation}
\frac{\partial \left(\overline{uw}_L\right)}{\partial z} = - u_{*0}^2 \left[ f'_m \cos(\delta) - f_m \delta' \sin(\delta) \right],
\label{EqdUWdz1}
\end{equation}
\begin{equation}
\frac{\partial \left(\overline{vw}_L\right)}{\partial z} = u_{*0}^2 \left[ f'_m \sin(\delta) + f_m \delta' \cos(\delta) \right].   
\label{EqdVWdz1}
\end{equation}
\end{subequations}
After differentiating one more time we arrive at: 
\begin{widetext}
\begin{subequations}
\begin{equation}
\frac{\partial^2 \left(\overline{uw}_L\right)}{\partial z^2} = - u_{*0}^2 \left[ f''_m \cos(\delta) - 2 f'_m \delta' \sin(\delta) - f_m \delta'' \sin(\delta) -f_m \delta'^2 \cos(\delta)\right], 
\label{Eqd2UWdz2}
\end{equation}
\begin{equation}
\frac{\partial^2 \left(\overline{vw}_L\right)}{\partial z^2} = u_{*0}^2 \left[ f''_m \sin(\delta) + 2 f'_m \delta' \cos(\delta) 
+ f_m \delta'' \cos(\delta) - f_m \delta'^2 \sin(\delta)\right].   
\label{Eqd2VWdz2}
\end{equation}
\end{subequations}
By combining Eqs.~\ref{EqEkmanCombined1}, \ref{EqUW2}, \ref{EqVW2}, \ref{Eqd2UWdz2}, \ref{Eqd2VWdz2}, and simplifying we get:
\begin{multline}
f_m f''_m \cos^2(\delta) - 2 f_m f'_m \delta' \sin(\delta)\cos(\delta) - f_m^2 \delta'' \sin(\delta)\cos(\delta) - f_m^2 \delta'^2 \cos^2(\delta) + \\
f_m f''_m \sin^2(\delta) + 2 f_m f'_m \delta' \sin(\delta)\cos(\delta) 
+ f_m^2 \delta'' \sin(\delta)\cos(\delta) - f_m^2 \delta'^2 \sin^2(\delta) = 0.
\label{EqEkmanCombined2}
\end{multline}
\end{widetext}
By invoking the Pythagorean trigonometric identity, we can further simplify this equation to: 
\begin{equation}
    f_m f''_m - f^2_m \delta'^2 = 0.
\label{EqEkmanCombined3}    
\end{equation}
\noindent Thus,
\begin{equation}
    \delta' = \sqrt{\frac{f''_m}{f_m}}.
\label{EqdDeltadz1}
\end{equation}
Substituting $f_m$ and $f_m''$ from Eqs.~\ref{ustar} and \ref{Eqd2fmdz2} in \ref{EqdDeltadz1}, we find:
\begin{equation}
    \delta' = \left(\frac{\sqrt{\alpha \left(\alpha -1 \right)}}{h}\right) \left(1-\frac{z}{h}\right)^{-1}.
\label{EqdDeltadz2}
\end{equation}
The second derivative of $\delta$ is: 
\begin{equation}
\delta'' = \frac{\sqrt{\alpha \left(\alpha -1 \right)}}{h^2} \left(1-\frac{z}{h}\right)^{-2}.
\label{Eqd2Deltadz2}
\end{equation}
Please note that these derivatives have real values if $\alpha$ is greater than one. 

Now, we can directly estimate the $K_M$ profile from Eqs.~\ref{EqEkman5}, \ref{EqVW2}, \ref{Eqd2UWdz2}, \ref{EqdDeltadz2}, and \ref{Eqd2Deltadz2}:
\begin{widetext}
\begin{subequations}
\begin{align}
    K_M & = - \frac{f_{cor} \overline{vw}_L} {\partial^2 (\overline{uw}_L)/ \partial z^2}\\
        & = + \frac{f_{cor} f_m \sin(\delta)}{f''_m \cos(\delta) - 2 f'_m \delta' \sin(\delta) - f_m \delta'' \sin(\delta) - f_m \delta'^2 \cos(\delta)}\\
        & = \frac{f_{cor} h^2 }{\left(2 \alpha - 1 \right)\sqrt{\alpha \left( \alpha - 1 \right)} }\left(1 - \frac{z}{h} \right)^2.
\end{align}
\label{Km}
\end{subequations}
\end{widetext}
We would like to emphasize that this formulation of $K_M$ is derived directly from the Ekman equations with very limited assumptions. To the best of our knowledge, a similar formulation and derivation have not been reported in the literature. 

Similar to other Ekman layer findings, Eq.~\ref{Km} is only valid in the outer layer. It does not represent surface layer conditions. In the literature, various patching and asymptotic matching approaches \citep[e.g.,][]{taylor15,blackadar68,brown74,zilitinkevich75} have been proposed to combine outer layer and inner layer (i.e., surface layer) solutions. A nice overview was given by \cite{hess02}. In Sects.~VI and VII, we utilize an unorthodox strategy. 

\section{Conventional K-profile Approach}

One of the most widely used first-order formulation for $K_M$ is the K-profile approach \citep{stensrud07}.  O'Brien~\cite{obri70}  was  one of  the  first  researchers  to propose  a K-profile  which  portrays  desirable  surface layer  behavior,  attains a maximum value within the planetary boundary layer (PBL), and decreases to a background diffusion level above the PBL. Based on a second-order closure model, Brost and Wyngaard~\cite{brost78} proposed a different K-profile formulation for stably stratified flows: 
\begin{equation}
        K_M = \frac{\left( \kappa z u_{*0} \right)}{\phi_M} \left(1 - \frac{z}{h} \right)^{p}.
        \label{Kprof}
\end{equation}
Here $\phi_M$ is a type of non-dimensional velocity gradient, defined later. The exponent $p$ is a-priori not known. For neutrally stratified flows in the surface layer, Eq.~\ref{Kprof} reduces to $K_M = \kappa z u_*$; this equation is in complete agreement with the well-established logarithmic law of the wall.  Furthermore,  for stably stratified surface layers,  one can deduce a stability-corrected logarithmic law of the wall \citep[e.g.,][]{businger71} from Eq.~\ref{Kprof}. 

The K-profile approach was modified by several researchers \cite{holtslag91,holtslag93,hong96,hong06,noh03,troen86} for its application in the unstable regime. They included a counter-gradient term to include the effects of large-scale eddies on local fluxes. They also considered entrainment fluxes at the top of the PBL.  

In the context of numerical stability,  the K-profile approach is quite robust \citep{beljaars92}. Thus, it is not surprising that it is widely used in numerical weather prediction models.  As a matter of fact,  the default PBL scheme (called the YSU scheme) of the popular Weather Research and Forecasting (WRF) model uses the K-profile approach for both unstable and stable conditions.  

In spite of its wide usage, Eq.~\ref{Kprof} suffers from two limitations.  First,  there is uncertainty in the value of the exponent $p$.  Brost and Wyngaard~\cite{brost78} found $p$ to be equal to 1.5 based on their simulations.  On the other hand, based on field campaign data from Minnesota,  Sorbjan~\cite{sorbjan89} found $p = 1$.  In the absence of reliable  field observations Troen and Mahrt~\cite{troen86} used an integer value of $p = 2$.  The local scaling hypothesis by Nieuwstadt~\cite{nieuwstadt84} also leads to $p = 2$.  In the present study, we estimate $p$ analytically. 

The other limitation is related to the parameterization of $\phi_M$.
The K-profile formulation uses the following normalized velocity gradient: 
\begin{subequations}
\begin{equation}
\phi_M = \left( \frac{\kappa z S}{u_{*0}} \right). 
\label{phiMS1}
\end{equation}
Please note that in this equation surface friction velocity ($u_{*0}$) is used.  However,  in contrast to well-known surface layer formulations,  $z$ is not limited to the depth of the surface layer.  Instead,  $z$ ranges from the surface to the top of the boundary layer.  $\phi_M$ is commonly parameterized as \cite{brost78}: 
\begin{equation}
\phi_M = 1 + c \left( \frac{z}{L} \right) = 1 + c \left( \frac{z}{h} \right)\left( \frac{h}{L} \right),
\label{phiMS2}
\end{equation}
\end{subequations}
where $c$ is a constant and often assumed equal to 5.  In the surface layer,  for $z/L < 1$,  numerous studies documented the validity of this equation.  However,  its applicability for the outer layer (i.e.,  above surface layer) is questionable.  Furthermore, this equation (incorrectly) implies that the logarithmic law of the wall applies to the entire boundary layer for neutral condition (i.e., $h/L = 0$).  

\section{Alternative K-profile Approaches}

In this section, we derive two competing K-profile formulations. Both the formulations are applicable for the entire SBL (i.e., including the surface layer and the outer layer). 

\subsection{Option 1}

Multiplying both sides of Eq.~\ref{EqKm3} by $\left(\kappa z/u_{*0}\right)$, we get:
\begin{subequations}
\begin{equation}
    \left( \frac{\kappa z}{u_{*0}} \right) u_{*L}^2 = K_M \left( \frac{\kappa z S}{u_{*0}} \right),
\end{equation}
By using Eq.~\ref{ustar}, we deduce: 
\begin{equation}
    \left( \kappa z u_{*0} \right) f_m = K_M \phi_M.
\end{equation}
Thus,
\begin{equation}
    K_M = \frac{\left( \kappa z u_{*0} \right) f_m}{\phi_M} = \frac{\left( \kappa z u_{*0} \right)}{\phi_M} \left(1 - \frac{z}{h} \right)^{\alpha}.
    \label{Kprof1}
\end{equation}
\end{subequations}
This equation is identical to the one proposed by Brost and Wyngaard~\cite{brost78} based on second-order modeling. Except, in case of Eq.~\ref{Kprof1}, the exponent $\alpha$ is the same as in Eq.~\ref{ustar}. 

\subsection{Option 2}

An alternate expression for $K_M$ profile can be found by multiplying Eq.~\ref{EqKm3} by $(\kappa z/u_{*L})$:
\begin{subequations}
\begin{equation}
    \left( \frac{\kappa z}{u_{*L}} \right) u_{*L}^2 = K_M \left( \frac{\kappa z S}{u_{*L}} \right),
\end{equation}
or, 
\begin{equation}
    \left( \kappa z u_{*L} \right) = K_M \phi_{ML}.
\end{equation}
By using Eq.~\ref{ustar}, we get:
\begin{equation}
    \left( \kappa z u_{*0} \right) f_m^{1/2} = K_M \phi_{ML},
\end{equation}
Hence,
\begin{equation}
    K_M = \frac{\left( \kappa z u_{*0} \right) f_m^{1/2}}{\phi_{ML}} = \frac{\left( \kappa z u_{*0} \right)}{\phi_{ML}} \left(1 - \frac{z}{h} \right)^{\alpha/2}.
    \label{Kprof2}
\end{equation}
\end{subequations}
In these equations, $\phi_{ML} (= \kappa z S/u_{*L})$ is a local non-dimensional velocity gradient as it utilizes local friction velocity ($u_{*L}$) from height $z$. It is straightforward to show that:
\begin{equation}
\phi_M = \phi_{ML} f_m^{1/2}. 
\label{phiMS3}
\end{equation}
Equation~\ref{phiMS3} implies that $\phi_M$ decreases more strongly with height than $\phi_{ML}$. Several past simulation studies \citep[e.g.,][]{basu06,zilitinkevich07,vandeWiel08} found the following parameterization for $\phi_{ML}$:
\begin{equation}
\phi_{ML} = 1 + c_L \left( \frac{z}{\Lambda} \right).
\label{phiML2}
\end{equation}
In those studies,  $c_L$ was found to be between 3 and 5.  By regression analysis of field observations,  Mahrt and Vickers~\cite{mahrt03} found $c_L$ = 3.7. Here,  we assume $c_L = 4$. 

\section{Matching of $K_M$ Profiles in the Outer Layer}

Thus far, we have derived 3 different K-profiles. Note that Eq.~\ref{Km} is only valid in the outer layer. In contrast, Eqs.~\ref{Kprof1} and \ref{Kprof2} are valid in the entire SBL. In the following sub-sections, we match these $K_M$ profiles for the outer layer when $z/L \gg 1$ and $z/\Lambda \gg 1$. 

\subsection{Option 1}

In the outer layer, for $z/L \gg 1$, Eq.~\ref{Kprof1} simplifies to:

\begin{subequations}
\begin{align}
    K_M & = \frac{\left( \kappa z u_{*0} \right) f_m}{c \left(\frac{z}{L}\right)},\\
        & = \left(\frac{\kappa}{c}\right) \left( u_{*0} L \right) f_m,\\
        & = \left(\frac{\kappa}{c}\right) \left( u_{*0} L\right) \left(1 - \frac{z}{h} \right)^{\alpha}.
\end{align}
\end{subequations}
If we assume $\alpha = 2$, then we get:
\begin{equation}
    K_M = \left(\frac{\kappa}{c}\right) \left( u_{*0} L \right) \left(1 - \frac{z}{h} \right)^2.
    \label{Km_match1}
\end{equation}
Similarly, by plugging in $\alpha = 2$ in Eq.~\ref{Km}, we get: 
\begin{equation}
    K_M = \frac{f_{cor} h^2}{3\sqrt{2}} \left(1 - \frac{z}{h} \right)^2.
    \label{Km_match2}
\end{equation}
By directly matching Eq.~\ref{Km_match1} with Eq.~\ref{Km_match2} we arrive at:
\begin{equation}
    \frac{f_{cor} h^2}{3\sqrt{2}}  =  \left(\frac{\kappa}{c}\right) \left( u_{*0} L \right).
\end{equation}
Further simplification leads to the SBL height parameterization of Zilitinkevich~\cite{zilitinkevich72}:
\begin{subequations}
\begin{equation}
    h = \gamma_1 \sqrt{\frac{u_{*0} L}{f_{cor}}},
\end{equation}
where
\begin{equation}
    \gamma_1 = \sqrt{3 \sqrt{2} \left(\frac{\kappa}{c}\right)}.
\end{equation}
\end{subequations}
For $c = 5$, we have: $\gamma_1 = 0.583$.

\subsection{Option 2}

Similar to option 1, in the outer layer, for $z/\Lambda \gg 1$, Eq.~\ref{Kprof2} simplifies to:

\begin{subequations}
\begin{align}
    K_M & = \frac{\left( \kappa z u_{*0} \right) f_m^{1/2}}{c_L \left(\frac{z}{\Lambda}\right)},\\
        & = \left(\frac{\kappa}{c_L}\right) \left( u_{*0} \Lambda \right) f_m^{1/2},\\
        & = \left(\frac{\kappa}{c_L}\right) \left( \frac{u_{*0} L f_m^{2}}{f_h} \right).
\end{align}
\end{subequations}
In this derivation, we used Eq.~\ref{Obukhov} in the conversion of $\Lambda$ to $L$. If we assume $\alpha = 3/2$ and $\beta = 1$, we get:
\begin{equation}
    K_M = \left(\frac{\kappa}{c_L}\right) \left( u_{*0} L \right) \left(1 - \frac{z}{h} \right)^2.
\end{equation}
Comparing this equation with Eq.~\ref{Km}, we arrive at:
\begin{subequations}
\begin{equation}
    h = \gamma_2 \sqrt{\frac{u_{*0} L}{f_{cor}}},
\end{equation}
where
\begin{equation}
    \gamma_2 = \sqrt{\sqrt{3} \left(\frac{\kappa}{c_L}\right)}.
\end{equation}
\end{subequations}
If $c_L = 4$, we get $\gamma_2 = 0.416$. 

Interestingly, both the options 1 and 2 lead to the SBL height parameterization of Zilitinkevich~\cite{zilitinkevich72}. The estimated proportionality constants, $\gamma_1$ and $\gamma_2$, fall within the range of previous observation-based and simulation-based empirical values. Furthermore, based on the literature, both $\alpha = 3/2$ and $\alpha = 2$ are plausible in SBLs. However, using $\phi_M = 1 + 5 z/L$ for the entire SBL, as commonly done in practice, does not seem physically meaningful and is in contrast with observations \citep[e.g.,][among many others]{holtslag84}. From this perspective, option 2 seems to be a better option.   

\section{Conclusion}

Fifty years ago, Zilitinkevich~\cite{zilitinkevich72} proposed a formulation for SBL height by using boundary layer scaling arguments. In this study, we derive the same formulation from an analytical approach involving the Ekman layer equations. In addition, we provide novel derivations for eddy viscosity profiles in the SBL. Our approach makes use of the following assumptions: (i) turbulent fluxes decrease monotonically with height and go to zero at the top of the boundary layer; (ii) the K theory is applicable for stably stratified conditions; and (iii) under steady-state condition, for flows over homogeneous and flat terrain, geostrophic balance holds in the boundary layer. None of these assumptions are unorthodox. In our future work, we hope to extend our analytical approach to derive geostrophic drag laws for SBLs.   

\begin{acknowledgements}
The first author is indebted to Pouriya Alinaghi, Branko Kosovi\'{c}, and Arqu\'{i}medes Ruiz-Columbi\'{e} for useful discussions and their assistance in cross-checking the analytical derivations. The second author acknowledges the many discussions and interactions he had with Sergej Zilitinkevich over the years.
\end{acknowledgements}

\bibliography{zilitinkevich}   

\end{document}